\documentclass{article}

\usepackage{graphicx}
\RequirePackage{url}
\usepackage{fancyvrb} 
\usepackage{amssymb,amsmath,amsfonts}
\usepackage{enumitem}
\usepackage{bm}
\usepackage{optidef}
\usepackage{empheq}

\usepackage{soul}

\usepackage[citecolor=red,colorlinks=true,linkcolor=blue,linktoc=page]{hyperref} 

\usepackage{arxiv}

\usepackage[utf8]{inputenc} 
\usepackage[T1]{fontenc}    
\usepackage{hyperref}       
\usepackage{url}            
\usepackage{booktabs}       
\usepackage{amsfonts}       
\usepackage{nicefrac}       
\usepackage{microtype}      
\usepackage{lipsum}
\usepackage{graphicx}
\graphicspath{ {./images/} }

\usepackage{graphicx}%
\usepackage{multirow}%
\usepackage{amsmath,amssymb,amsfonts}%
\usepackage{amsthm}%
\usepackage{mathrsfs}%
\usepackage[title]{appendix}%
\usepackage{xcolor}%
\usepackage{textcomp}%
\usepackage{manyfoot}%
\usepackage{booktabs}%
\usepackage{algorithm}%
\usepackage{algorithmicx}%
\usepackage{algpseudocode}%
\usepackage{listings}%
\usepackage{bm,float,placeins}

\newcommand{\p}[2]{\frac{\partial #1}{\partial #2}}

\theoremstyle{thmstyleone}%
\theoremstyle{thmstyletwo}%

\theoremstyle{thmstylethree}%

\title{Conserved Quantities of Optimal Continuous-Thrust Trajectories in A Central Gravitational Field}

\author{
 Aimar Negrete \\
  Department of Aerospace Engineering\\
  Iowa State University\\
  Ames, Iowa 50011 \\
  \texttt{aimar@iastate.edu} 
  \And
 Ossama Abdelkhalik \\
  Department of Aerospace Engineering\\
  Iowa State University\\
  Ames, Iowa 50011 \\
  \texttt{ossama@iastate.edu} \\
}

\begin{document}
\maketitle
\begin{abstract}
This paper presents a mathematical derivation for three new conserved quantities in the motion of spacecraft on optimal continuous-thrust trajectories in a central gravitational field. The process presented in this paper is rooted in Noether’s theorem that connects the point symmetries of a dynamic system with the associated conservation laws of the system. 
In the approach presented in this paper, the system's Lagrangian is modified to account for the non-conservative control force. Using this generalized Lagrangian, the action functional to be minimized is written. Then, Killing equations are formulated to find the dynamic symmetries for this system. In this paper a process is laid out for how to solve the Killing equations{; Noether's theorem is applied to this solution of the Killing equations to write the conserved quantities of the system.}. 
Conserved quantities are presented for both the two-dimensional and the three-dimensional trajectories in several different coordinate frames. Numerical simulations and mathematical proofs are used to demonstrate that the computed quantities are conserved.
\end{abstract}

\section{Introduction}

Optimal control theory addresses the problem of determining control functions that will cause a dynamic system to optimize a certain performance criterion while satisfying differential equation constraints. While the formal establishment of optimal control theory occurred in the latter half of the 20$^{\mathrm{th}}$ century, its foundational roots can be traced to the early twentieth century when calculus of variations experienced rigorous formalization influenced by developments in real analysis and functional analysis. 
It can be stated that feedback control systems emerged prominently in the engineering efforts of WWII. The development of radar systems, and guidance technologies spurred significant advancements in applied control theory. Engineers such as Hendrik Bode and Harry Nyquist at Bell Labs developed tools such as the Bode plot and Nyquist criterion, which analyzed the frequency response and stability margins of feedback systems.
These techniques informed the subsequent formal development of control laws optimizing a cost functional subject to dynamic constraints. It was this collaboration between engineers and mathematicians that laid the groundwork for systems theory \cite{wiener1949}. In fact, investments in aerospace and automation early in the latter half of the twentieth century drove the demand for optimal guidance and control systems \cite{bellman1957,pontryagin1962}.
During that period, Bellman's principle of optimality emerged which states that ``an optimal policy has the property that, whatever the initial state and initial decision are, the remaining decisions must constitute an optimal policy with regard to the state resulting from the first decision,'' \cite{bellman1957}. Then, Pontryagin introduced a boundary-value formulation for optimal control problems \cite{pontryagin1962}. Suppose a system
\begin{equation}\label{PMP01}
\dot{\bm{x}}(t) = f(\bm x(t), \bm u(t)),
\end{equation}
{has initial conditions $\bm{x}(t_0)=\bm{x}_0$} and performance index
\begin{equation}\label{PMP02}
J = \int_{t_0}^{t_f} L(\bm x(t), \bm u(t), t)\, dt,
\end{equation}
{with continuous constraints on the control $\bm{g}(\bm u)\leq \bm 0$. Additionally, assume that the state vector $\bm{x}$ is continuous and differentiable.}
The maximum principle states that there exists a co-state vector $\bm\lambda(t)$ such that the Hamiltonian
\begin{equation}\label{PMP03}
H = \bm\lambda^T \bm f(\bm x, \bm u) + L(\bm x, \bm u, t)
\end{equation}
is maximized by the optimal control $\bm u^*(t)$:
\begin{equation}\label{PMP04}
H(\bm x^*, \bm u^*, \bm \lambda, t) = \max_{\bm u} H(\bm x, \bm u, \bm \lambda, t).
\end{equation}
and that the adjoint variable $\lambda(t)$ satisfies:
\begin{equation}\label{PMP05}
\dot{\bm\lambda} = -\frac{\partial H}{\partial \bm x}
\end{equation}
{There is no requirement that the control vector $\bm{u}(t)$ be differentiable, or even continuous. }

Driven by the space race, and advancements in digital computing, the above foundational work was followed by two decades of rapid expansion in the theory and applications of optimal control. The era saw widespread application of optimal control in aerospace (trajectory optimization for Apollo missions), economics (optimal growth models), and industrial automation. Bryson and Ho’s textbook \textit{Applied Optimal Control} \cite{Bryson1975} became a standard reference, disseminating the theory to engineering audiences. Extensions such as time-varying, constrained, and stochastic optimal control were actively developed, laying the groundwork for modern robust and adaptive control methods.

Methods to solve optimal control problems are typically classified into direct and indirect approaches. 
Indirect methods are grounded in the calculus of variations and the Pontryagin's Maximum Principle (PMP) presented in Eqs.~\eqref{PMP01}-\eqref{PMP05}.
Indirect methods derive and solve the first-order necessary conditions for optimality, usually formulated as a boundary value problem (BVP). 
The solution of the resulting boundary value problem is typically attempted using shooting methods \cite{betts1998survey,stoer2002introduction}. {Higher order conditions (such as the Legendre-Clebsch condition) may be included in the solution of the optimal control problem as well.}
{Indirect methods offer high accuracy, provide better theoretical guarantees, and can offer insight into the structure of the problem;} they have often been applied in the aerospace trajectory optimization problem as in \cite{betts2010practical}. Yet indirect methods often suffer from numerical sensitivity and convergence issues. In particular, they require a good initial guess (sometimes for the co-states which could be hard to find.) Direct methods, on the other hand, {are also very important branch of optimal control. Generally, direct methods} convert the optimal control problem directly into a nonlinear programming problem and are often more robust \cite{rao2009survey} {than indirect methods. } 

The present authors recently investigated an alternate approach for optimal control, based on a generalized Lagrangian of the dynamic system, as detailed in \cite{OCP2025AAShawaii}. This new optimal control approach eliminates the need for the adjoint variable (costates).  This paper presents new development on this new optimal control approach. Specifically this paper presents a method to derive conserved quantities on the optimal trajectory of a dynamic systems, and applies this method to continuous-thrust optimal space trajectories in a central gravitational field. 
This paper is organized as follows. Section~\ref{Background} presents a necessary background in Lie Theory and Noether’s Theorem, the mathematical foundation for the present work. Section~\ref{GL} summarizes the new generalized Lagrangian approach in optimal control. Section~\ref{ConservQ} presents the development of the conservative quantities for continuous-thrust trajectories. {The conserved quantities are developed for fixed time, minimum energy trajectories with an unbounded control magnitude.} Finally Section~\ref{NumericalExample} presents numerical case studies which demonstrate the time invariance of the derived conserved quantities in the continuous thrust optimal control problem.  

\section{Background}\label{Background}
\subsection{Lie Theory and Noether's Theorem}
In the development of this work, we make extensive use of the Lie theory of one parameter groups. In particular, we use a theorem from Noether \cite{Noether1971} which was developed through an application of Lie theory to the problem of conservation laws in dynamical systems. A brief summary of both of these topics is presented in this section.

\subsubsection{Lie Theory} \label{bglt}
The Lie theory of one parameter groups was introduced by Sophus Lie in 1893 in order to characterize infinitesimal transformations. Specifically, Lie was focused on the continuous symmetries of differential equations to arrive at solutions to these equations. A symmetry of a differential equation is an infinitesimal transformation which leaves the differential equation invariant. Symmetries transform solutions of differential equations into other solutions. For an ordinary differential equation (ODE), invariance under a one-parameter symmetry group allows the order of the ODE to be reduced by one \cite{olver1993applications}. If we can arrive at the ODE through a variational principle (as is the case with the generalized Lagrangian approach), the order of the problem can be reduced by two. The search for symmetries of a system can thus provide a tangible benefit in solving the differential equations. The summary in this section follows the development of Ref. \cite{cohen1911introduction}, which can be consulted for further detail.

To begin, consider a system defined by two coordinates $\bm{x}=[t,y]^T$. Let a set of transformations be defined on $\bm{x}$ by a parameter $a$ such that the transformed coordinates $\bm{x}'=[t', y']$ are given by 
\begin{equation}
    \begin{split}
        t'&=\phi(t,y,a) \\
        y'&= \psi(t,y,a)
    \end{split}
    \label{oneParameterGroup}
\end{equation}
\noindent An example of a transformation can be a translation of $t$ by an amount $c\cdot a$. That is, 
\begin{equation}
    \begin{split}
        t' &= t + c\cdot a \\ 
        y' &= y
    \end{split}
    \label{translation}
\end{equation}
\noindent where $c$ is some constant. Without loss of generality, let $\phi$ and $\psi$ be the identity transformation when $a = a_0$. That is, {$t=\phi(t,y,a_0)$ and $y=\psi(t,y,a_0)$}. Suppose now that we take an infinitesimal transformation of our coordinates around the identity transformation at $a_0$ by an {infinitesimal} parameter $\varepsilon\ll1$.
\begin{equation}
    \begin{split}
        t' &= \phi(t,y,a_0+\varepsilon) \\ 
        y' &= \psi(t,y, a_0+\varepsilon) 
    \end{split}
\end{equation}
\noindent Expanding this in a Taylor series about $\varepsilon=0$, 
\begin{equation}
    \begin{split}
        t' &= \phi(t,y,a_0) + \left(\p{\phi}{a}\right)_{a_0}\varepsilon + \mathcal{O}(\varepsilon^2) = t+\xi(t,y)\varepsilon+\mathcal{O}(\varepsilon^2) \\
        y' &= \psi(t,y,a_0) + \left(\p{\psi}{a}\right)_{a_0}\varepsilon + \mathcal{O}(\varepsilon^2) = y+\eta(t,y)\varepsilon+\mathcal{O}(\varepsilon^2) \\
    \end{split}
    \label{groupTaylor}
\end{equation}
\noindent The (continuous) functions $\xi(t,y)$ and $\eta(t,y)$ define the infinitesimal transformation. The \textit{generator} of the infinitesimal transformation is defined \cite{cohen1911introduction} as 
\begin{equation}
    G = \xi(t,y)\p{}{t} + \eta(t,y)\p{}{y}
    \label{groupGenerator}
\end{equation}
\noindent This generator describes, to first order, how the coordinates $\bm{x}$ change with respect to the infinitesimal transformation induced by $\xi$ and $\eta$. The generator of the infinitesimal transformation for the example of the translation in Eq. \eqref{translation} is $G=c\p{}{x}$. As mentioned previously, Lie theory is concerned with finding invariants of the group action. Any (local) invariant $\zeta(t,y)$ of our group will be a solution to the partial differential equation
\begin{equation}
    G\{\zeta\}=\xi(t,y)\p{\zeta}{t} + \eta(t,y)\p{\zeta}{y} = 0
\end{equation}
\noindent This defines a characteristic system of ordinary differential equations which can be integrated to arrive at the invariances.  For our purposes, we will be looking at the invariances of the generalized Lagrangian we give in Section \ref{GL}. In this functional, we have derivatives of the state variables, and so one needs to know how the derivatives will also change with respect to the infinitesimal transformation of the coordinates.

Until now, the relationship between $t$ and $y$ has been left unstated. For the application into differential equations which is required, define $t$ as the independent variable and $y$ as a dependent variable. There can be as many independent and dependent variables as necessary (see Ref. \cite{olver1993applications}) but for this section we maintain one independent variable and one dependent variable. The group generator $G$ describes the change in coordinates to first order in $\varepsilon$ under the infinitesimal transformation described by $\eta$ and $\xi$. As mentioned above, it is necessary also to know how the derivatives of the coordinates will change as a function of $\eta$ and $\xi$. To this end, the notion of extended group generators is introduced. The generator of the first extended group is defined as 
\begin{equation}
    E = \xi(t,y)\p{}{t} + \eta(t,y)\p{}{y} + \tilde\eta(t,y,\dot{y})\p{}{\dot{y}} = G + \tilde{\eta}(t,y,\dot{y})\p{}{\dot{y}}
\end{equation}

\noindent where $\dot{y}=dy/dt$. It can be shown through a Taylor expansion as in Eq. \eqref{groupTaylor} that $\tilde{\eta}=\dot{\eta}-\dot{y}\dot{\xi}$, such that the generator of the first extended group can be written as 
\begin{equation}
    E = \xi(t,y)\p{}{t} + \eta(t,y)\p{}{y} + \left(\dot{\eta}(t,y)-\dot{y}\dot{\xi}(t,y)\right)\p{}{\dot{y}} 
\end{equation}
\noindent The operator $E$ describes to first order in $\varepsilon$ the change in a first order differential equation with respect to the transformations induced by $\xi(t,y)$ and $\eta(t,y)$. As was the case of the group generator $G$, the condition for a function $\chi(t,y,\dot{y})$ to be invariant under the extended group $E$ is that $E\{\chi\}=0$. 

For the Cartesian system described in Section \ref{GL}, the independent variable is the time $t$, and the dependent variables are the position vector $\bm{r}$ and the control vector $\bm{u}$; hence the vector of generalized coordinates (dependent variables) is $\bm{q}=[x,y,z,u_x,u_y,u_z]^T$. With this definition, the generator of the first extended group for the optimal control system is given by 
\begin{equation}
    E(t,\bm{q},\dot{\bm{q}}) = \xi(t,\bm{q})\p{}{t} + \sum_{i=1}^n \eta_i(t,\bm{q}) \p{}{q_i} + \left[ \dot \eta_i - \dot\xi\dot q_i\right]\p{}{\dot{q}_i}
    \label{extendedgroupgenerator}
\end{equation}
\noindent where the total time derivatives of the transformation's generators can be written in terms of partial derivatives as
\begin{equation}
    \begin{split}
        \dot\xi &= \p{\xi}{t}+\sum_{k=1}^n \p{\xi}{q_k}\dot{q}_k \\
        \dot\eta_i  &=\p{\eta_i}{t} + \sum_{k=1}^n \p{ \eta_i}{ q_k}\dot{q}_k
    \end{split}
\end{equation}
\subsubsection{Noether's Theorem}
Noether's theorem provides the mathematical connection between the dynamic symmetries of a system and associated conservation laws of the dynamic system \cite{noether1983invariante,Noether1971}. Prior to Noether, the primary way of finding conserved quantities of a system came directly from the Euler-Lagrange equations. Consider for example an $n$ degree of freedom physical system with Lagrangian function is $L(t,\bm{q},\dot{\bm q})$. By applying the Euler-Lagrange equations to $L$, we arrive at the equations of motion of the system. The Euler-Lagrange equations state that for $i = 1,2,...,n$,
\begin{equation}
    \frac{d}{dt}\left(\p{L}{\dot{x}_i}\right)-\p{L}{x_i}=\psi_i(L)=0
    \label{eulerLagrange}
\end{equation}
\noindent Straight away, these equations show that if $\partial L/\partial x_i=0$, then $d/dt\left(\partial L / \partial \dot{x}_i \right)=0$, and so $\partial L / \partial\dot{x}_i=\mathrm{const.}$. Noether's theorem is a generalization of this idea in the terms of Lie theory. 

To apply Noether's theorem to a particular system, it is necessary that the equations of motion be derivable from a variational principle; this is the case with the generalized Lagrangian approach outlined in Section \ref{GL}. Noether's theorem searches for invariances of the variational principle from which the equations of motion are derived. This is a departure from the classical Lie theory, which looks for invariances of the differential equations themselves. Consider an arbitrary functional $$\Gamma = \int_{t_0}^{t_f}L(t,\bm{q},\dot{\bm{q}})\,dt$$ Additionally, following in the form of Eq. \eqref{groupTaylor}, we define an infinitesimal transformation for each coordinate (including the time). 
\begin{equation}
    \begin{split}
        t' &= t + \xi(t,\bm{q})\varepsilon + \mathcal{O}(\varepsilon^2) \\ 
        \bm{q}' &= \bm{q} + \bm{\eta}(t,\bm{q})\varepsilon + \mathcal{O}(\varepsilon^2)
    \end{split}
    \label{timeCoordTransform}
\end{equation}

\noindent {The functions $\xi$ and $\bm{\eta}$ are defined in the same way as in Eq. \eqref{groupTaylor} }. Here, $\bm{\eta}$ is an $n\times 1$ vector of transformation generators, each one corresponding to a coordinate. Note that this remains a one parameter family of transformations in $\varepsilon\ll 1$. Following the definition in Ref. \cite{neuenschwander}, the functional $\Gamma$ is said to be invariant under the transformations in Eq. \eqref{timeCoordTransform} if  
\begin{equation}
    \left|\int_a^b L(t,\bm{q},\dot{\bm{q}}) dt - \int_a^b L(t',\bm{q}',\dot{\bm{q}}') dt'\right| \sim  \varepsilon^s
    \label{NoetherTherm01}
\end{equation}

\noindent where $s\geq 2$. This criterion simply states that, if invariant, the functional evaluated with the original coordinates and the functional evaluated in the new coordinates will evaluate to the same value down to the linear term of $\varepsilon$. Noether's theorem states that if the functional $\Gamma$ is both stationary and invariant under the infinitesimal transformation generated by $\xi$ and $\bm{\eta}$, then 
\begin{equation}
\Phi=\frac{\partial L}{\partial \dot{\bm{q}}}(\bm{\eta}-\xi\dot{\bm{q}}) + L\xi\equiv\mathrm{const}.
    \label{NoetherTheorem}
\end{equation}

\noindent When the Euler-Lagrange equations \eqref{eulerLagrange} are applied to a functional to arrive at the equations of motion, the functional is being rendered stationary. Therefore, the conserved quantity described in Eq. \eqref{NoetherTheorem} will be a conservation law along the trajectories of the system. 

There is a more general form of Noether's theorem which we will apply to the continuous thrust optimal control problem, which comes from a property of the variational principles from which the equations of motion of a system can be written. Suppose that a system's dynamics can be described by a Lagrangian $L(t,\bm{q},\dot{\bm{q}})$. {Let $\varphi(t,\bm{q})$ be a differentiable function of only time and the generalized coordinates (not of the generalized velocities). Then, the Lagrangian $\tilde{L}=L+\frac{d\varphi(t,\bm{q})}{dt}$ will yield the same equations of motion as $L$ when rendered stationary.} With this additional liberty, the so-called "divergence invariant" \cite{neuenschwander} form of Noether's theorem can be stated.
\begin{equation}
    \Phi = \frac{\partial L}{\partial \dot{\bm{q}}}(\xi \dot{\bm{q}} - \bm{\eta}) - \xi L + \varphi(t,\bm{q}) \equiv \mathrm{const.}
    \label{noetherDivergence}
\end{equation}

\noindent {The proofs of both the classical form (Eq. \eqref{NoetherTheorem}) and the divergence invariant form (Eq. \eqref{noetherDivergence}) of Noether's theorem may be found in Ref. \cite{neuenschwander}.} The divergence invariant form of Noether's theorem will be applied to find the additional conserved quantities of our dynamic system to allow for additional solutions that may not be found when using the classical form of Noether's theorem. 

\section{Generalized Lagrangian}\label{GL}
We continue with a brief overview of a method of optimal control which the present authors developed to formulate and solve optimal control problems without the need for costates in the solution process. The development of the generalized Lagrangian method is investigated in more detail in Ref. \cite{OCP2025AAShawaii}. The generalized Lagrangian is an extension of the variational methods of analytical mechanics applied to optimal control problems. 

There are many distinct methods of analytical mechanics which have been developed over the centuries; our analysis is based off of Hamilton's Law of Varying Action. In his law, Hamilton defined the action functional of a conservative dynamical system as the difference between the system's total kinetic energy, $T$, and the total potential energy, $V$.
\begin{equation}
    S=\int_{t_0}^{t_f}(T-V)\,dt
\end{equation}

\noindent Hamilton proved that the system's equation of motion will minimize $S$. That is to say, a state trajectory $\bm{q}(t),\dot{\bm{q}}(t)$ which satisfies the equations of motion ($f(t,\bm{q},\dot{\bm{q}})=0$) will have a lower value for $S$ than any neighboring path $\bar{\bm{q}},\dot{\bar{\bm{q}}}$. More formally, $S(t,\bm{q},\dot{\bm{q}}) \leq S(t,\bar{\bm{q}},\dot{\bar{\bm{q}}})$ will hold for all neighboring paths. Strictly speaking, it is not mathematically necessary that $S$ be a minimum in the general case, only that it be stationary; however, as $T$ is inherently unbounded for physical systems ($T\rightarrow \infty$ as $||\dot{\bm{q}}||\rightarrow\infty$), the action functional has no maximum. There may be a minimum or a saddle point, which may be determined if desired through the second variation of the action functional. 

If a nonconservative work is to be found in the dynamical system, Hamilton's law can be generalized to include this term in the action.
\begin{equation}
    S = \int_{t_0}^{t_f}(T-V+W_{nc})\,dt
\end{equation}

\noindent where $W_{nc}=\int_{\bm{q}_0}^{\bm{q}_f}\bm{F}\cdot\,d\bm{q}$ is the nonconservative work done by {a generalized} force $\bm{F}$. 

The generalized Lagrangian approach derives from the same idea as Hamilton's principle, with the difference lying in the choice of generalized coordinates and of the action functional. In the new approach, we begin with the assertion that, for a given optimal control problem, there exists a functional, $\tilde{S}=\int 
\tilde{L}\,dt$, which will be minimized by the state and control trajectories of the optimally controlled system. $\tilde{L}$ is the eponymous generalized Lagrangian alluded to above. In the new approach, the original generalized coordinates, $\bm{q}$, as well as the control variables, $\bm{u}$ are treated as a new set of modified generalized coordinates, $\tilde{\bm{q}}$. The new action functional will now be minimized by the equations of motion for $\tilde{\bm{q}}$. These equations of motion include those for the original state vector, as well as differential equations for the optimal control. 

In this paper, we focus our efforts on finding conserved quantities in the continuous thrust two body problem. Specifically, we focus our search on the conserved quantities of the minimum energy optimal control problem with cost functional to be minimized
\begin{equation}
    \mathrm{min}\,\mathcal{J}=\bm{\psi}(t_f,\bm{x}(t_f)) + \frac{1}{2}\int_{t_0}^{t_f}\bm{u}^T\bm{u}\,dt
\end{equation}
{where $\bm{\psi}$ includes any boundary terms to be be minimized or boundary conditions to be satisfied.}

To this end, we will present two such generalized Lagrangian functions for the two body problem {which are valid for the minimum energy optimal control problem}; one in Cartesian coordinates, and one in spherical coordinates. The method for arriving at these generalized Lagrangians is not the focus of this paper; a more in-depth exploration of this technique can be found in Ref. \cite{OCP2025AAShawaii};  appendix A further presents a brief proof that this generalized Lagrangian yields the correct equations of motion and also equations for the optimal control.

In Cartesian coordinates, the generalized Lagrangian can be written as 
\begin{equation}
    L = \dot x \dot u_x + \dot y \dot u_y + \dot z \dot u_z - \frac{\mu}{r^3}\left(xu_x + yu_y + zu_z \right) + \frac12 \left(u_x^2 + u_y^2 + u_z^2\right)
    \label{cartesiangeneralizedLagrangian}
\end{equation}

\noindent We may find the differential equations which minimize Eq. \eqref{cartesiangeneralizedLagrangian} by computing the first variation of $\tilde{S}=\int L\,dt$ with respect to $\tilde{\bm{q}}=[x,y,z,u_x,u_y,u_z]^T$ and requiring that it vanish. For example, taking the variation with respect to $u_x$,
\begin{equation}
\begin{split}
    \frac{\delta \tilde S}{\delta u_x}&=\int_{t_0}^{t_f} \left( \p{L}{x}\delta u_x + \p{L}{\dot{u}_x}\delta \dot u_x \right)\,dt \\ 
    &= \int_{t_0}^{t_f} \left( \p{L}{u_x} - \frac{d}{dt} \p{L}{\dot{u}_x}\right)\,\delta  u_x dt\\
    &= \int_{t_0}^{t_f} \left( -\frac{\mu}{r^3}x + u_x - \frac{d}{dt} \dot x \right)\,\delta  u_x dt\\
    \end{split}
\end{equation}

\noindent Integration by parts was used to rewrite the integral in terms of $\delta u_x$; the boundary terms were dropped in keeping with the general assumption of Hamilton's principle that variations vanish at the endpoints. Asserting that the variational derivative with respect to $u_x$ vanish for all $\delta u_x$ yields 
\begin{equation}
    \ddot x = -\frac{\mu}{r^3}x + u_x
\end{equation}

\noindent {To arrive at the remaining state and control differential equations, the variational derivative must be taken with respect to $y,z,u_x,u_y,$ and $u_z$ as was done above with $x$} Applying the same approach for the remaining generalized coordinates yields the full set of state and control differential equations. 
\begin{equation}
    \begin{split}
        \ddot{\bm{r}} &= -\frac{\mu}{r^3}\bm{r} + \bm{u} \\ 
        \ddot{u}_x &= -\frac{\mu}{r^3}u_x + \frac{3\mu}{r^5}\left(x^2u_x + xyu_y + xzu_z \right) \\ 
        \ddot{u}_y &= -\frac{\mu}{r^3}u_y + \frac{3\mu}{r^5}\left( y^2u_y + xyu_x + yzu_z\right) \\
        \ddot{u}_z  &= -\frac{\mu}{r^3}u_z + \frac{3\mu}{r^5}\left(z^2u_z + xzu_x + yzu_y \right)
    \end{split}
    \label{cartesianEOMs}
\end{equation}
{Note that these differential equations describe the optimal evolution of the control for an unbounded control magnitude. That is to say, there is no continuous constraint on the control direction or magnitude. }

The generalized Lagrangian can also be formulated for a two body problem with continuous thrust in spherical coordinates. For the development of the conserved quantities in Section IV, we will be using the Cartesian generalized Lagrangian from Eq. \eqref{cartesiangeneralizedLagrangian}. Afterwards, we will demonstrate how to transform the conserved quantities from the Cartesian two body problem to the spherical two body problem. Let $\bm{q}_s=[r,\theta,\phi]^T$ be the radius and two polar angles such that the Cartesian position vector of the spacecraft is given by $$\bm{r}=r\left[ \begin{array}{c} \cos{\phi}\cos{\theta} \\ \cos{\phi}\sin{\theta} \\ \sin{\phi} \end{array}\right]$$ The generalized Lagrangian for this problem is given by 
\begin{equation}
\begin{split}
        L_s = \dot{r} \dot{u}_r+\dot{\theta} \dot{u}_{\theta}+\dot{\phi} \dot{u}_{\phi}-\frac{\mu}{r^2}u_r+\left( r{\dot{\phi}}^2 + r {\dot{\theta}}^2 \cos^2{\phi} \right)u_{r}-\frac{1}{2}{\dot{\theta}}^2\sin{(2\phi)}u_\phi- \\
        \frac{2\dot{r}}{r}\left( \dot{\phi}  u_{\phi } + \dot{\theta} u_{\theta }\right) + 2 \dot{\phi} \dot{\theta} \tan{(\phi)}u_{\theta }  +\frac{1}{2}\left(u_r^2 + u_\theta^2 + u_\phi^2\right)
\end{split}
    \label{sphericalgeneralizedLagrangian}
\end{equation}
\noindent The equations of motion and control derived from the spherical generalized Lagrangian are presented in the appendix. One can transform between the Cartesian coordinates to the polar coordinates through the following transformations. 
\begin{equation}
    \begin{split}
        r &= \sqrt{x^2 +y^2+z^2} \\ 
        \theta &= \tan^{-1}\left(\frac{y}{x}\right) \\ 
        \phi &= \sin^{-1}\left(\frac{z}{r}\right) \\ 
        u_r &= \cos\phi\cos\theta \,u_x + \cos\phi\sin\theta\,u_y + \sin\phi\,u_z \\ 
        u_\theta &= -\sin\theta\,u_x + \cos\theta\,u_y \\ 
        u_\phi &= -\sin\phi\cos\theta\,u_x - \sin\phi\sin\theta\,u_y + \cos\phi \,u_z
    \end{split}
    \label{cartesian2spherical}
\end{equation}
\noindent The transformations are invertible and well defined everywhere where $r > 0$. These transformations will be utilized in Section \ref{CTS} to demonstrate how conserved quantities in one optimal control problem can be transformed into conserved quantities of an equivalent optimal control problem in alternative coordinates. 

\section{Conserved Quantities  for Continuous-Thrust Trajectories}\label{ConservQ}
In this section, we present a method to apply Lie theory to the optimal control action principle in order to find conserved quantities with the divergence invariant form of Noether's theorem. The classical form can be applied for a simpler set of PDEs by letting $\varphi=0$, at the cost of some freedom in the solution. {In the classical Hamilton's principle, the action of a conservative system is the time integral of the system's Lagrangian, $T-V$. For the optimal control procedure presented here, the action of the system is given by the time integral of the generalized Lagrangian (Eq. \eqref{cartesiangeneralizedLagrangian} for a Cartesian system)}. For the three dimensional continuous thrust optimal control problem, the action to be minimized is 
\begin{equation}
    S_{3D} = \int_{t_0}^{t_f}\left( \dot x \dot u_x + \dot y \dot u_y + \dot z \dot u_z - \frac{\mu}{r^3}\left(xu_x + yu_y + zu_z \right) + \frac12 \left(u_x^2 + u_y^2 + u_z^2\right)\right)\,dt
    \label{3daction}
\end{equation}

\noindent Although we will present the results for the general three dimensional case, we will instead elaborate the details of the two dimensional case, which is a subset of the three dimensional case. The action principle for the two dimensional case is given by 
\begin{equation}
    S_{2D} = \int_{t_0}^{t_f}L\,dt=\int_{t_0}^{t_f}\left( \dot x \dot u_x + \dot y \dot u_y  - \frac{\mu}{r^3}\left(xu_x + yu_y \right) + \frac12 \left(u_x^2 + u_y^2 \right)\right)\,dt
    \label{2daction}
\end{equation}
{which is arrived at by asserting $z,\dot{z},u_z,$ and $\dot{u}_z$ all vanish in Eq. \eqref{3daction}.} With a variational principle available from which the equations of motion can be derived, the search can now begin for symmetry generators of {$S_{2D}$}. The generators are arrived at through the application of a modified form of the invariance identity presented in section \ref{bglt}.

\subsection{Killing Equations}
The control vector $\bm{u}$ is treated as part of the generalized coordinates, hence the complete set of generalized coordinates for the two dimensional case is $\bm{q}^T=[x,y,u_x,u_y]=[\bm{r}^T, \bm{u}^T]$. The set of generators for the generalized coordinate transformation is $\bm{\eta}^T = [\eta_1,\eta_2,\eta_3,\eta_4]$, corresponding to the position vector transformation generators ($\eta_{1},\eta_2$) and the control vector transformation generators ($\eta_{3},\eta_4$).

We are looking for infinitesimal generators $\xi(t,\bm{q})$ and $\bm{\eta}(t,\bm{q})$ which will render the action integral invariant under an infinitesimal transformation. Once these functions are found, they can be applied to Noether's theorem to find the conserved quantities which we are after.  To find the generators which will satisfy this condition, we need to apply the generator of the first extended group to the action principle. Specifically, we need to solve for the generators which will satisfy 
\begin{equation}
    E\{L\} = -\dot{\xi}L + \dot{\varphi}
    \label{invarianceCriterion}
\end{equation}
{The derivation for this condition on the generators and $\varphi$ can be found in Ref. \cite{lutzky1978symmetry}.} The right hand side of the equation does not vanish due to the use of the divergence invariant form of Noether's theorem. Applying the operator $E$ to the Lagrangian in Eq, \eqref{2daction} will result in a polynomial in the generalized velocities $\dot{\bm{q}}=[\dot x, \dot y, \dot{u}_x, \dot u_y]^T$. Similarly, the right hand side of Eq. \eqref{invarianceCriterion} can be expanded in terms of partial derivatives to be expressed in terms of these generalized velocities through the expansion of the total time derivatives of $\xi$ and $\varphi$. 
\begin{equation}
     E\{L\}=-\dot\xi L + \dot\varphi = -\left(\p{\xi}{t} + \p{\xi}{x}\dot x + \p{\xi}{y}\dot y + \p{\xi}{u_x}\dot{u}_x + \p{\xi}{u_y}\dot{u}_y\right)L + \left(\p{\varphi}{t} + \p{\varphi}{x}\dot x + \p{\varphi}{y}\dot y + \p{\varphi}{u_x}\dot{u}_x + \p{\varphi}{u_y}\dot{u}_y\right)
\end{equation}
The result of this is a single scalar partial differential equation (PDE), which must be satisfied by $\xi(t,\bm{q})$ and $\bm{\eta}(t,\bm{q})$. A set of PDEs can be found from this single scalar equation by recognizing that, in the classical Noether's theorem, the generators $\xi(t,\bm{q})$ and $\bm{\eta}(t,\bm{q})$ are only allowed to be functions of time and generalized position, not generalized velocity. Because of this, the only way for Eq. \eqref{invarianceCriterion} to be satisfied is if the coefficients of all the velocity-dependent terms separately vanish. To illustrate this, suppose that applying $E$ to $L$ and moving all terms of Eq. \eqref{invarianceCriterion} to one side results in an equation such as 
\begin{equation}
    A(t,\bm{q},\xi,\bm{\eta}) + B(t,\bm{q},\xi,\bm{\eta})\dot{x} + C(t,\bm{q},\xi,\bm{\eta})\dot{x}\dot{u}_x + D(t,\bm{q},\xi,\bm{\eta})\dot{y}^2\dot{u}_y +...=0
\end{equation}
\noindent Then, we must require that the functions $A,B,C,D,...$ all separately vanish, leading to a system of PDEs. Once again, the reason we can group terms in this way is because the generators are not functions of the generalized velocities, and so the coefficients of the generalized velocity terms must all vanish separately. This set of partial differential equations is known as the Killing equations \cite{vujanovic1970group}, and the solution of these equations represents an invariance of the action functional induced by the generators of the infinitesimal transformations. The complete set of Killing equations for the 2D Cartesian continuous thrust optimal control problem is given in Eqs. \eqref{killingXi}-\eqref{killingPhi5}. 

\begin{table}[hbt!]
\begin{tabular}{ll}

\parbox{7cm}{
\begin{equation}
  \p{\xi}{x} = \p{\xi}{y} = \p{\xi}{u_x}=\p{\xi}{u_y} =0 
  \label{killingXi}
\end{equation}}

& 
\parbox{7cm}{
\begin{equation}
  \p{\eta_1}{u_x} =\p{\eta_2}{u_y}=\p{\eta_3}{x}=\p{\eta_4}{y} = 0 
  \label{killingEta}
\end{equation}}       

\\

\parbox{7cm}{
\begin{equation}
  \p{\eta_3}{u_x} + \p{\eta_1}{x} = \p{\xi}{t} 
  \label{killingXit1}
\end{equation}   }          

& 
\parbox{7cm}{
\begin{equation}
  \p{\eta_4}{u_y} + \p{\eta_2}{y} = \p{\xi}{t}
  \label{killingXit2}
\end{equation}  }     

\\

\parbox{7cm}{
\begin{equation}
  \p{\eta_1}{u_y} + \p{\eta_2}{u_x} = 0
  \label{killingEta12}
\end{equation}}

& 
\parbox{7cm}{
\begin{equation}
  \p{\eta_3}{y} + \p{\eta_4}{x} = 0
  \label{killingEta34}
\end{equation}   }         

\\

\parbox{7cm}{
\begin{equation}
  \p{\eta_1}{y} + \p{\eta_4}{u_x}= 0
  \label{killingEta14}
\end{equation} }       

& 
\parbox{7cm}{
\begin{equation}
  \p{\eta_2}{x} + \p{\eta_3}{u_y} =0 
  \label{killingEta23}
\end{equation}  }     

\\

\parbox{7cm}{
\begin{equation}
  \p{\eta_1}{t} = \p{\varphi}{u_x}
  \label{killingPhi1}
\end{equation}}        

& 
\parbox{7cm}{
\begin{equation}
  \p{\eta_2}{t}   = \p{\varphi}{u_y}
  \label{killingPhi2}
\end{equation} }      

\\

\parbox{7cm}{
\begin{equation}
  \p{\eta_3}{t} = \p{\varphi}{x}
  \label{killingPhi3}
\end{equation}}        

& 
\parbox{7cm}{
\begin{equation}
  \p{\eta_4}{t} = \p{\varphi}{y}
  \label{killingPhi4}
\end{equation} }      

\\

\multicolumn{2}{l}
{
\parbox{0.95\textwidth}{
\begin{equation}
    \begin{split}
        \left( \frac{3\mu x}{r^5}(xu_x + yu_y) - \frac{\mu}{r^3}u_x\right)\eta_1 + \left( \frac{3\mu y}{r^5}(xu_x + yu_y) - \frac{\mu}{r^3}u_y\right)\eta_2 + \left(u_x-\frac{\mu}{r^3}x\right)\eta_3 + \left( u_y-\frac{\mu}{r^3}y\right)&\eta_4 \\ 
        = \p{\varphi}{t} + \left(\frac{\mu}{r^3}(xu_x + yu_y) - \frac{1}{2}\left(u_x^2 + u_y^2\right)\right)&\p{\xi}{t}
    \end{split}
    \label{killingPhi5}
\end{equation} }
}
\end{tabular}
\end{table}

\noindent Equation \eqref{killingXi} is a condition on $\xi$ that it should only be a function of time. Accordingly, we let $\p{\xi}{t}\equiv\dot{\xi}=f(t)$, and replace the right hand sides of Eqs. \eqref{killingXit1} and \eqref{killingXit2} with $f(t)$. Similarly, the conditions in Eq. \eqref{killingEta} describe which variables each coordinate transformation is not a function of. The generator $\eta_1$ is not a function of $u_x$, $\eta_2$ is not a function of $u_y$, and so on. The rest of the PDEs are relations which must be satisfied by the transformation generators and their derivatives, and the next sections are dedicated to finding solutions of this set of PDEs and the associated conserved quantities that arise from these transformations. 

\subsection{Solving the Killing Equations}\label{SKE}
 We begin by examining Eq. \eqref{killingXit1}. From Eq. \eqref{killingEta}, we have that $\eta_1=\eta_1(t,y,u_x,u_y)$ and that $\eta_3=\eta_3(t,x,y,u_y)$. The right hand side of Eq. \eqref{killingXit1} is a strict function of time, as required by Eq. \eqref{killingXi}. For this to be satisfied, it must hold that
\begin{equation}
    \begin{split}
        \p{\eta_1}{x} &= C(t)-B(t,y,u_y) \\ 
        \p{\eta_3}{u_x} &= A(t) + B(t,y,u_y)
    \end{split}
    \label{eta1eta3xi}
\end{equation}
\noindent such that the sum of these two partial derivatives will be a function of only time. Additionally, it is required that $A(t)+C(t)=f(t)=\dot{\xi}$. Eq. \eqref{eta1eta3xi} can now be integrated with respect to $x$ for the first equation and $u_x$ for the second equation to arrive at 
\begin{equation}
    \begin{split}
        \eta_1 &= \big( C(t)-B(t,y,u_y)\big)x + D(t,y,u_y) \\ 
        \eta_3 &= \big( A(t) + B(t,y,u_y)\big)u_x + E(t,y,u_y)
    \end{split}
    \label{eta1eta3step1}
\end{equation}
\noindent A similar argument is applied to Eq. \eqref{killingXit2} to get 
\begin{equation}
    \begin{split}
        \eta_2 &= \big(H(t) - G(t,x,u_x)\big)y + I(t,x,u_x) \\ 
        \eta_4 &= \big(F(t) + G(t,x,u_x)\big)u_y + J(t,x,u_x) 
    \end{split}
    \label{eta2eta4step1}
\end{equation}
\noindent From Eqs. \eqref{eta1eta3step1} and \eqref{eta2eta4step1}, we see that $\eta_1$ is linear in $x$, $\eta_2$ is linear in $y$, $\eta_3$ is linear in $u_x$, and $\eta_4$ is linear in $u_y$. Substituting these forms for $\eta_3$ and $\eta_4$ into Eq. \eqref{killingEta34} yields
\begin{equation}
    \p{G(t,x,u_x)}{x}u_y+\p{J(t,x,u_x)}{x}+\p{B(t,y,u_y)}{y}u_x  +\p{E(t,y,u_y)}{y}=0
    \label{eta3eta4step2}
\end{equation}
\noindent If $G$ is not linear in $x$, then $\partial G/\partial x$ will be a function of $t,x,$ and $u_x$, and there will be a $(\partial G/\partial x)u_y$ term. However, $u_y$ only shows up elsewhere in the equation in functions of $t,y,u_x$, and $u_y$. Therefore, to require that Eq. \eqref{eta3eta4step2} vanish is to require that $G$ be at most linear in $x$, and (through a similar argument) that $B$ be at most linear in $y$. An additional requirement that arises from this is that $J$ be linear in $x$, and $E$ linear in $y$. The same argument can be applied in Eq. \eqref{killingEta12} to assert that $I$ must be linear in both $x$ and $u_x$, $N$ must be linear in $u_y$, $D$ must be linear in both $y$ and $u_y$, and that finally $L$ is linear in $u_x$.  The updated forms for $\eta_{1-4}$ can subsequently be substituted into Eqs. \eqref{killingEta14} and \eqref{killingEta23} and similar linearity conditions observed to arrive at the next intermediate form for the coordinate transformation generators which satisfies Eqs. \eqref{killingXi} to \eqref{killingEta23}. 
\begin{equation}
    \begin{split}
        \eta_1 &= c_1(t)x+c_2(t)y + c_3(t)u_y + h_1(t) \\ 
        \eta_2 &= c_4(t)x + c_5(t)y - c_3(t)u_x + h_2(t) \\ 
        \eta_3 &= c_6(t)y + (f(t)-c_1(t))u_x- c_4(t)u_y + h_3(t) \\ 
        \eta_4 &= -c_6(t)x-c_2(t)u_x+(f(t)-c_5(t))u_y + h_4(t)
    \end{split}
    \label{eta1234intermediate}
\end{equation}
\noindent The divergence term, $\varphi(t,x,y,u_x,u_y)$ can now be put together consistent with the coordinate transformation generators in Eq. \eqref{eta1234intermediate}. From Eq. \eqref{killingPhi1}, 
\begin{equation}
    \dot c_1x+\dot c_2y +  \dot c_3 u_y + \dot h_1 =\p{\varphi}{u_x}
\end{equation}
\noindent Integrating with respect to $u_x$, it can be said that $\varphi = (\dot c_1x+\dot c_2y +  \dot c_3 u_y + \dot h_1)u_x + p_1(t,x,y,u_y)$, where $p_1$ is a constant of integration with respect to $u_x$. Substitution of this expression for $\varphi$ into Eq. \eqref{killingPhi2} yields
\begin{equation}
    \begin{split}
        \dot c_4x + \dot c_5y - \dot c_3 u_x + \dot h_2 &= \dot c_3 u_x + \p{p_1}{u_y} \\
        \implies\p{p_1(t,x,y,u_y)}{u_y} &= \dot c_4x + \dot c_5y - 2\dot c_3 u_x + \dot h_2
    \end{split}
\end{equation}
\noindent The only way for the left and right hand sides to be equal is for $\dot c_3=0$, which means that $c_3(t)=k_3$ for some constant $k_3$. Integration with respect to $u_y$ provides the form $p_1(t,x,y,u_y)=(\dot c_4x+\dot c_5y+\dot h_2)u_y+p_2(t,x,y)$. Once again, the updated expression for $\varphi$ can be passed in to Eq. \eqref{killingPhi3} for the condition that 
\begin{equation}
\begin{split}
    \dot c_6 y + (\dot f-\dot c_1)u_x-\dot c_4 u_y+\dot h_3 &= \dot c_1u_x + \dot c_4 u_y + \p{p_2}{x} \\ 
    \implies \p{p_2(t,x,y)}{x} &= \dot c_6 y + (\dot f -2\dot c_1)u_x - 2\dot c_4 u_y + \dot h_3
\end{split}
\end{equation}
\noindent As before, the only way that the two sides can be equal is if the coefficients of the $u_x$ and $u_y$ terms vanish, as there is no way for the left hand side to contain any terms in $u_x$ or $u_y$. Therefore, $\dot f=2\dot c_1$ and $c_4(t)=k_4$, thus yielding the expression $p_2(t,x,y)=(\dot c_6y+\dot h_3)x + p_3(t,y)$. We can take the updated form for $\varphi$ into Eq. \eqref{killingPhi4} for the requirement that 
\begin{equation}
    \begin{split}
        -\dot c_6 x - \dot c_2 u_x + (\dot f - \dot c_5)u_y + \dot h_4 &= \dot c_2 u_x + \dot c_5 u_y + \dot c_6 x + \p{p_3}{y} \\ 
        \implies -2\dot c_6 x - 2\dot c_2 u_x + (\dot f - 2\dot c_5)u_y + \dot h_4 &= \p{p_3(t,y)}{y}
    \end{split}
\end{equation}
\noindent Following the same logic as before, $c_2(t)=k_2$, $c_6(t)=k_6$, and $\dot f = 2\dot c_5$. This gives $p_3(t,y)=\dot h_4 y + p_4(t)$, and
\begin{equation}
    \varphi(t,x,y,u_x,u_y) = \frac12\dot f(t) x u_x + \frac12\dot f(t)yu_y + \dot h_1(t) u_x + \dot h_2(t) u_y + \dot h_3(t) x + \dot h_4(t) y + p_4(t)
    \label{phiIntermediate}
\end{equation}
\noindent The only remaining work left to be done to find the generators of the invariances is to evaluate Eq. \eqref{killingPhi5} and solve for the coefficients in $\xi, \bm{\eta},$ and $\varphi$. For notational brevity, define $d=xu_x+yu_y$. Substituting Eqs. \eqref{eta1234intermediate} and \eqref{phiIntermediate} into Eq. \eqref{killingPhi5} and collecting terms yields
\begin{equation}
\begin{split}
        \frac{3\mu}{r^5}d\left(c_1x^2+k_2xy+k3xu_y+h_1x+k_4xy+c_5y^2-k_3yu_x+h_2y\right)-\frac{\mu}{r^3}\left(\left(c_1+\frac{1}{2}f\right)d + h_3 x + h_4 y\right) \\ 
        +k_6(yu_x-xu_y)+\frac{1}{2}(u_x^2+u_y^2) = \frac{1}{2}\ddot f d + \ddot h_1u_x + \ddot h_2 u_y + \ddot h_3 x + \ddot h_4 y + \dot p_4 - \frac{1}{2}f(u_x^2+u_y^2)+\frac{\mu}{r^3}df
\end{split}
\end{equation}
\noindent It can be shown that the only consistent solution to this equation is for $c_1(t)=c_2(t)=h_1(t)=h_2(t)=h_3(t)=h_4(t)=f(t)=k_3=k_6=\dot{p}_4(t)=0$, and $k_2=-k_4$. With all the Killing equations satisfied, the generators for the time and coordinate transformations can be written as 
\begin{equation}
    \begin{split}
        \eta_1 &= k_2 y \\ 
        \eta_2 &= -k_2 x \\ 
        \eta_3 &= k_2 u_y \\ 
        \eta_4 &= k_2 u_x \\ 
        \xi &= k_\xi \\ 
        \varphi &= k_\varphi
    \end{split}
\end{equation}
\noindent There are three constants, so there are three linearly independent generators which can be found by letting in turn each constant equal 1 and all others equal to zero. However, the conserved quantity associated with $\varphi=k_\varphi$ is trivial ($\Phi_\varphi=k_\varphi$) , so it can be disregarded. The generators for the other two symmetries are $\bm{\eta}_1=[y, -x, u_y, -u_x]^T$, $\xi_1=0$ and $\bm{\eta}_2 = \bm{0}, \xi_2=1$. 

Now that the Killing equations have been solved, we can provide some insight into the transformations themselves and thus the type of conservation law which will arise from the application of each. Consider first the transformation in time and $\bm{q}$ induced by $\bm{\eta}_2$ and $\xi_2$. This set of generators states that the action of the system will be unchanged under a constant transformation of the system time. Classically, time invariance such as this will lead to conservation of the system's total energy; thus we will describe this as an "energy-type" conservation law. The infinitesimal transformation given by $\bm{\eta}_1$ for $x$ and $y$ is the infinitesimal transformation for a rotation of $\bm{r}$ by an angle $\varepsilon$. Similarly, the transformation components for $u_x$ and $u_y$ describe a rotation of $\bm{u}$ by the same angle $\varepsilon$. Rotational invariance in the classical sense will lead to a conservation of angular momentum, and so we will describe the associated conserved quantity as an "angular momentum"-type conservation law.

\subsection{Conserved Quantities}\label{TCQ}
The three dimensional Killing equations, presented in the appendix, can be solved in much the same way as the two dimensional case presented above. Through the procedure, we find that there are four sets of linearly independent generators which satisfy the Killing equations in three dimensions, and therefore four linearly independent conserved quantities. In the two dimensional case, two of these conserved quantities identically vanish, and we are reduced to the two conservation laws solved for above. For the 3 dimensional case, the "angular momentum"-type transformations have $\xi=0$ and
\begin{equation}
\begin{split}
    \bm{\eta}_1 &= [y, -x, 0, u_y, -u_x, 0]^T \\ 
    \bm{\eta}_2 &= [0, z, -y, 0, u_z, -u_y]^T \\ 
    \bm{\eta}_3 &= [-z, 0, x, -u_z, 0, u_x]^T
\end{split}
\label{eta123}
\end{equation}
\noindent For the "energy"-type transformation, we have $\bm{\eta}_4=\bm{0}$ and $\xi=1$. For all transformations, we have $\varphi=0$. Applying each set of $\bm{\eta},\xi$ to Noether's theorem \eqref{noetherDivergence} provides the following conserved quantities. 
\begin{equation}
\begin{split}
    \Phi_1 &= x\dot{u}_y - \dot{x}u_y - y\dot{u}_x + \dot{y}u_x \\ 
    \Phi_2 &= y\dot{u}_z - \dot{y}u_z - z\dot{u}_y + \dot{z}u_y \\ 
    \Phi_3 &= z\dot{u}_x - \dot{z}u_x - x\dot{u}_z + \dot{x}u_z \\ 
    \Phi_4 &= \dot{x}\dot{u}_x + \dot{y}\dot{u}_y + \dot{z}\dot{u}_z +\frac{\mu(xu_x+yu_y+zu_z)}{\sqrt{x^2+y^2+z^2}^3} - \frac{1}{2}(u_x^2+u_y^2+u_z^2)
\end{split}
\label{conservationLaws3D}
\end{equation}

The conserved quantity $\Phi_4$ is equivalent to the requirement that the Hamiltonian in the classical minimum energy optimal control problem be conserved. The other three, however, are new conserved quantities in this problem to the best of the authors' knowledge. In addition to the implicit proof provided by the application of Noether's theorem, the system and control equations of motion can be used to directly prove the invariance of these quantities. Consider $\Phi_3$ for example, and take the total time derivative of this function.
\begin{equation}
\begin{split}
    \dot{\Phi}_3 &= \dot z \dot u_x + z\ddot u_x - \ddot z u_x - \dot z \dot u_x - \dot x \dot u_z - x \ddot u_z + \ddot x u _z + \dot x \dot u_z \\
            &= z\ddot u_x - \ddot z u_x - x \ddot u_z + \ddot x u _z  \\
            &= 
            \begin{array}[t]{l}
                 z\left(-\frac{\mu}{r^3}u_x + \frac{3\mu}{r^5}\left(x^2u_x + xyu_y + xzu_z \right)\right) - u_x\left(-\frac{\mu}{r^3}z + u_z\right) \\
                - x\left( -\frac{\mu}{r^3}u_z + \frac{3\mu}{r^5}\left(z^2u_z + xzu_x + yzu_y \right)\right) + u_z\left(-\frac{\mu}{r^3}x + u_x\right)\\ 
            \end{array}\\
            &= \frac{3\mu z}{r^5}\left(x^2u_x + xyu_y + xzu_z \right) - u_xu_z - \frac{3\mu x}{r^5}\left(z^2u_z + xzu_x + yzu_y \right) + u_xu_z \\
            &= \frac{3\mu}{r^5}\left( (x^2zu_x + xyzu_y + xz^2u_z) - (xz^2u_z + x^2zu_x + xyzu_y) \right) \\ 
            &= 0
\end{split}
\end{equation}
\noindent The state and control equations of motion from Eq. \eqref{cartesianEOMs} were substituted into the expression in the third step, and some algebra shows that the time derivative of $\Phi_3$ vanishes. Similar proofs can be constructed for both $\Phi_1$ and $\Phi_2$, demonstrating that these quantities are invariants on the solution curves of the system. This will be further shown numerically in Section IV for several optimal continuous thrust transfers.

\subsection{Coordinate Transformation of Conserved Quantities}\label{CTS}
For some problems, it may be more convenient to solve the optimal control problem in spherical coordinates, cylindrical coordinates, or in general some other alternative coordinate set that may be arrived at through a coordinate transformation of the Cartesian coordinates. These include classical orbital elements, equinoctial elements, and so on. In these cases, it is possible to transform the conserved quantities derived above into the new coordinate systems, and have a corresponding conserved quantity in the new coordinates. The only requirement is that the generalized Lagrangian be available for the new set of coordinates, as it is required for the application of Noether's theorem. 

Simply taking the original conserved quantities in the Cartesian frame and transforming them is not sufficient to yield a conserved quantity in the new frame. Instead, we will transform the infinitesimal generators which yielded $\Phi_i$ into the new coordinates, then take those new generators back into Noether's theorem with the new Lagrangian to get the conserved quantities. It is important to note than in this section, we will be discussing two different types of coordinate transformations. The first transformation is the finite coordinate transformation between the Cartesian frame and the spherical frame, defined by the functions in Eq. \eqref{cartesian2spherical}. The second type of transformation is the infinitesimal transformation which we have previously described in Section III, from which Noether's theorem is developed. These are distinct coordinate transformations, and they should not be confused. 

To demonstrate the technique, consider first a planar version of the generalized Lagrangian presented in Eq. \eqref{sphericalgeneralizedLagrangian}. In this case, $\phi=\dot{\phi}=u_\phi=\dot{u}_\phi=0$, and the generalized Lagrangian reduces to 
\begin{equation}
    L_P = \dot{r} \dot{u}_r+\dot{\theta} \dot{u}_{\theta}-\frac{\mu}{r^2}u_r+  r {\dot{\theta}}^2u_{r}-\frac{2\dot{r}\dot{\theta} }{r}u_{\theta }  +\frac{1}{2}\left(u_r^2 + u_\theta^2\right)
    \label{polargeneralizedLagrangian}
\end{equation}
\noindent It was shown in Section \ref{SKE} that an angular momentum-like conserved quantity in the planar two body problem is characterized by the transformation $\bm{\eta}_{2D}=[-y,x,-u_y,u_x]^T,\,\, \xi_{2D}=0$. We would like to find what transformation in the polar two body problem would lead to an equivalent conserved quantity in the new coordinates. From the definition of the infinitesimal transformation in Eq. \eqref{groupTaylor}, the infinitesimally transformed Cartesian coordinates are $\bm{q}'_C=\bm{q_c} + \varepsilon\cdot \bm{\eta}_{2D}$, or 
\begin{equation}
    \begin{split}
        x' &= x - \varepsilon y \\ 
        y' &= y + \varepsilon x \\ 
        u_x' &= u_x - \varepsilon u_y \\ 
        u_y' &= u_y + \varepsilon u_x \\ 
    \end{split}
    \label{newCartesianCoords2d}
\end{equation}
\noindent The finite transformation between the Cartesian frame and the polar frame is given by 
\begin{equation}
    \begin{split}
        r &= \sqrt{x^2 + y^2}\\
        \theta &= \tan^{-1}{\frac{y}{x}} \\
        u_r &= u_x\cos{\theta} + u_y\sin{\theta}  \\ 
        u_\theta &= u_y\cos{\theta}-u_x\sin{\theta}
    \end{split}
\end{equation}
\noindent Note that the control variables are included in both the finite and infinitesimal coordinate transformations, because they are part of the expanded set of generalized coordinates in which we are working. We can now take these infinitesimally transformed coordinates and perform the finite transformation to bring them into the polar frame. The end goal with this transformation is to see what the equivalent one parameter family of transformations is in the polar coordinates which will have the same effect on the polar coordinates as $\bm{\eta}_{2D}$ has on the Cartesian coordinates. The transformed coordinates are 
\begin{equation}
    \begin{split}
        r' &= \sqrt{(x-\varepsilon y)^2 + (y+\varepsilon x)^2}\\
        \theta' &= \tan^{-1}{\left(\frac{y+\varepsilon y}{x-\varepsilon y}\right)} \\
        u_r' &= (u_x-\varepsilon u_y)\cos{\theta'} + (u_y + \varepsilon u_x)\sin{\theta'}  \\ 
        u_\theta '&= (u_y + \varepsilon u_x)\cos{\theta'}-(u_x-\varepsilon u_y) \sin{\theta'}
    \end{split}
\end{equation}
The one parameter family of transformations induce by $\bm{\eta}_{2D}$ is an infinitesimal transformation in $\varepsilon$, so only the first order effect of the infinitesimal transformation on the polar coordinates needs to be considered. To this end, we take a Taylor series of Eq. \eqref{newSpherical1} around $\varepsilon=0$.
\begin{equation}
    \begin{split}
        r' &= r + \frac{1}{2}r\varepsilon^2 + \mathcal{O}(\varepsilon^3) \\ 
        \theta' &= \theta + \varepsilon + \mathcal{O}(\varepsilon^3) \\ 
        u_r' &= u_r+\frac{1}{2}u_r\varepsilon^2 + \mathcal{O}(\varepsilon^3) \\ 
        u_\theta ' &= u_\theta + \frac{1}{2}u_\theta \varepsilon^2 +\mathcal{O}(\varepsilon^3) \\ 
    \end{split}
    \label{infinitesimalPolar}
\end{equation}
\noindent In Eq. \eqref{groupTaylor}, the generator of the infinitesimal coordinate transformation was defined to be the linear term of the Taylor series in $\varepsilon$ around $\varepsilon=0$. Comparing this to the infinitesimally transformed polar coordinates in Eq. \eqref{infinitesimalPolar}, it {can be seen} that {the invariance with respect to the polar angle $\theta$ is} $\bm{\eta}_p=[0, 1, 0, 0]^T$ by finding the linear terms in $\varepsilon$ of the Taylor expansion. {This} is the infinitesimal transformation in polar coordinates which corresponds to $\bm{\eta}=[-x,y,-u_x,u_y]^T$ in Cartesian coordinates. {Transforming $\bm{\eta}$ from Cartesian to polar coordinates in this way allows one to avoid having to go through the Killing equations in polar coordinates to find the infinitesimal transformation which will generate invariances in the new set of coordinates.} Noether's theorem can now be applied to the new infinitesimal transformation in the polar coordinates (with the polar generalized Lagrangian in Eq. \eqref{polargeneralizedLagrangian}) to find the associated conservation law in the polar two body problem. 
\begin{equation}
    \begin{split}
        \Phi_p &= \p{L_p}{\dot {\tilde{\bm{q}}}}(\bm{\eta}_p - \xi\dot{\tilde{\bm{q}}}) \\ 
              &= \left[\begin{array}{cccc} \dot{u}_r -\frac{2\dot{\theta}}{r}u_\theta & \dot{u}_\theta +2r\dot{\theta}u_r-2\frac{\dot{r}}{r}u_\theta& \dot{r} & \dot{\theta} \end{array}\right]\left[ \begin{array}{c}0 \\ 1 \\ 0 \\ 0 \end{array}\right] \\ 
              &= \dot{u}_\theta + 2r\dot{\theta}u_r - \frac{2\dot{r}}{r}u_\theta
    \end{split}
    \label{phip}
\end{equation}
\noindent {As the coordinate transformation between Cartesian and polar coordinates did not also involve a time transformation, we can leave $\xi=0$ in the new application of Noether's theorem.} It can be verified that $\Phi_p$ is a conserved quantity in the polar two body problem in the same way as it was demonstrated in Section \ref{TCQ} for the Cartesian conserved quantities. Taking the derivative of $\Phi_p$,
\begin{equation}
    \begin{split}
        \dot{\Phi}_p &= \ddot u_\theta +2(\dot r\dot \theta +r\ddot \theta)u_r + 2r\dot\theta\dot u_r +\frac{2}{r^2}(\dot r^2 -\ddot r)u_\theta - \frac{2\dot r}{r}\dot u_\theta \\ 
        &= \left(-2(\dot r\dot \theta +r\ddot \theta)u_r - \frac{2}{r^2}(\dot r^2 -\ddot r)u_\theta + \frac{2\dot r}{r}\dot u_\theta - 2r\dot\theta\dot u_r\right) + \\
        &+\,\,2(\dot r\dot \theta +r\ddot \theta)u_r + 2r\dot\theta\dot u_r +\frac{2}{r^2}(\dot r^2 -\ddot r)u_\theta - \frac{2\dot r}{r}\dot u_\theta \\ 
        &= 0
    \end{split}
\end{equation}
\noindent The equation of motion for $\ddot u_\theta$ was substituted in from the appendix, letting all $\phi$ terms be zero. As its time derivative vanishes, $\Phi_p$ is a conserved quantity for the minimum energy optimal control problem solved in polar coordinates.

Consider now the 3-dimensional optimal control problem in spherical coordinates characterized by the generalized Lagrangian in Eq. \eqref{sphericalgeneralizedLagrangian}, with coordinate transformations given by Eq. \eqref{cartesian2spherical}. To transform the conserved quantities in Eq. \eqref{conservationLaws3D} to this new coordinate frame, we begin with the set of infinitesimal generators $\bm{\eta}_1=[y,-x,0,u_y,-u_x,0]^T, \,\xi_1=0$ from which we originally derived $\Phi_1$ in the Cartesian frame. As before, the transformed coordinates can be written as 
\begin{equation}
    \begin{split}
        x' &= x + \varepsilon y \\ 
        y' &= y - \varepsilon x \\ 
        z' &= z \\
        u_x' &= u_x + \varepsilon u_y \\ 
        u_y' &= u_y - \varepsilon u_x \\ 
        u_z' &= u_z
    \end{split}
    \label{newCartesianCoords1}
\end{equation}
\noindent Placing these infinitesimal transformations into the finite coordinate transformations which map from the Cartesian frame to the spherical frame yields a set of infinitesimally transformed spherical coordinates. 
\begin{equation}
    \begin{split}
        r' &= \sqrt{(x+\varepsilon y)^2 + (y - \varepsilon x)^2 + z^2} \\
        \theta' &= \tan^{-1}\left(\frac{y - \varepsilon x}{x + \varepsilon y} \right) \\
        \phi' &= \sin^{-1}\left(\frac{z}{r'}\right) \\ 
        u_r' &= -\cos\phi'\cos\theta' \,(u_x+\varepsilon u_y) + \cos\phi'\sin\theta'\,(u_y-\varepsilon u_x) + \sin\phi'\,u_z\\
        u_\theta' &= -\sin\theta'\,(u_x+\varepsilon u_y) + \cos\theta'\,(u_y-\varepsilon u_x) \\
        u_\phi' &= -\sin\phi'\cos\theta'\,(u_x+\varepsilon u_y) - \sin\phi'\sin\theta'\,(u_y-\varepsilon u_x) + \cos\phi' \,u_z        
    \end{split}
    \label{newSpherical1}
\end{equation}
\noindent We proceed as before, taking the Taylor series of Eq. \eqref{newSpherical1} around $\varepsilon=0$ and neglecting terms on the order of $\varepsilon^2$ or higher to see the equivalent infinitesimal transformation in spherical coordinates. 
\begin{equation}
    \begin{split}
        r' &= r + \mathcal{O}(\varepsilon^2) \\
        \theta' &= \theta -\varepsilon+ \mathcal{O}(\varepsilon^2)\\
        \phi'&= \phi + \mathcal{O}(\varepsilon^2)\\
        u_r' &= u_r + \mathcal{O}(\varepsilon^2)\\
        u_\theta' &= u_\theta+ \mathcal{O}(\varepsilon^2)\\
        u_\phi' &= u_\phi + \mathcal{O}(\varepsilon^2)
    \end{split}
\end{equation}
\noindent The corresponding generator for the conserved quantity in the spherical frame, like in the polar frame, is a simple rotation about $\theta$; that is $\bm{\eta}_{1,s}=[0,1,0,0,0,0]^T$. 

This process can be repeated for the other two generators of the conserved quantities in the Cartesian frame to get the associated conserved quantities in the spherical frame. For $\bm{\eta_2}=[0, z, -y, 0, u_z, -u_y]^T $, the equivalent infinitesimal transformation in the spherical frame is given by $\bm{\eta}_{2,s}=[0, \cos{\theta}\tan{\phi},-\sin{\theta},\sin{\phi}u_y, \cos{\theta}\sec{\phi}u_\phi, -\cos{\theta}\sec{\phi}u_\theta]^T$, where $u_y$ can be expressed in terms of the spherical coordinates from Eq. \eqref{cartesian2spherical}. Similarly, the equivalent infinitesimal transformation for $\bm{\eta}_3= [-z, 0, x, -u_z, 0, u_x]^T$ in the spherical frame is $\bm{\eta}_{3,s}=[0, \sin{\theta}\tan{\phi},\cos{\theta},-\sin{\theta}u_x, \sin{\theta}\sec{\phi}u_\phi,\sin{\theta}\sec{\phi}u_\theta]^T$. The generators $\bm{\eta}_{1-3,s}$ can be applied to Noether's theorem to arrive at the corresponding conserved quantities in the spherical frame. 

This method is equally valid for transformation of the conserved quantities into any desired frame, as long as there exists an invertible differentiable mapping between the two sets of coordinates. It is necessary that the mapping be differentiable due to the Taylor series that is taken in finding the generators in the new coordinates. Since only the linear term of the series is required, it only needs to be once differentiable and not necessarily smooth.

\section{Numerical Demonstration}\label{NumericalExample}

To demonstrate the invariance of the presented conservation laws on optimal spacecraft trajectories, we consider two forms of a LEO to GEO transfer, both with similar initial and final conditions on the state variables. At the initial time, the spacecraft starts at a $6678$ km circular orbit around the Earth with an inclination of $28^\circ$, and at the final time the spacecraft is to be at a $42164$ km circular geostationary orbit with zero inclination. The first transfer is chosen with a short transfer time, and the second with a much more realistic time of flight for a continuous thrust spacecraft. The governing state and control differential equations are those given in Eq. \eqref{cartesianEOMs}. {For both transfers, the equations of motion were propagated with canonical units with $1 \,\mathrm{DU}=6378.137\,\mathrm{km}$ and $1\,\mathrm{TU}=806\,\mathrm{s}$. The initial conditions for the short time of flight transfer are given in Table \ref{ICs1}.}

\begin{table}[hbt!]
\caption{\label{ICs1} Initial conditions for short transfer in canonical units, $t_f=183.444\,\mathrm{TU}$}
\centering
\begin{tabular}{crcr}
\hline
State & Initial Value & Control & Initial Value \\\hline
$x$ & $1.047$         & $u_x$       & $2.97849\cdot 10^{-5}$ \\
$y$ & $0$             & $u_y$       & $8.02439\cdot 10^{-3}$\\ 
$z$ & $0$             & $u_z$       & $2.90147\cdot 10^{-3}$\\ 
$\dot{x}$ & $0$       & $\dot{u}_x$ & $-7.96989\cdot10^{-3}$ \\ 
$\dot{y}$ & $0.86290$ & $\dot{u}_y$ & $1.62862\cdot 10^{-5}$ \\
$\dot{z}$ & $0.45881$ & $\dot{u}_z$ & $-1.09144\cdot 10^{-5}$ \\
\hline
\end{tabular}
\end{table}

\begin{figure}[hbt!]
    \centering
    \includegraphics[width=1\textwidth]{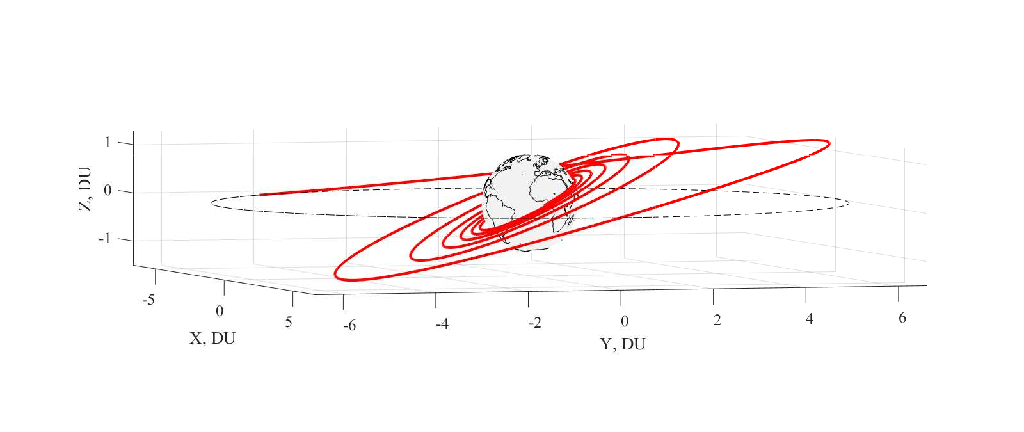}
    \caption{Optimal LEO to GEO transfer, $\Delta t\approx$ 40 hours}
    \label{shorttrajectory}
\end{figure}

\begin{figure}[hbt!]
    \centering
    \includegraphics[width=0.9\textwidth]{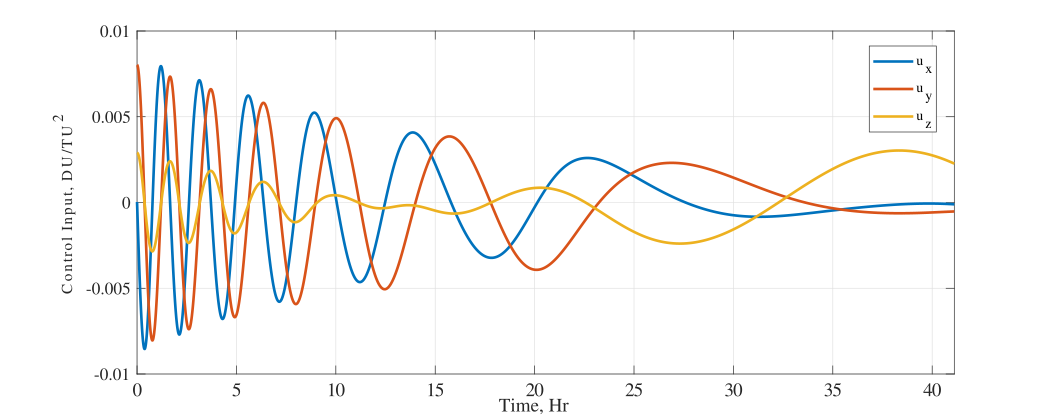}
    \caption{Control time history for the short time LEO to GEO transfer}
    \label{shorttrajectorycontrol}
\end{figure}

\begin{figure}[hbt!]
    \centering
    \includegraphics[width=0.8\textwidth]{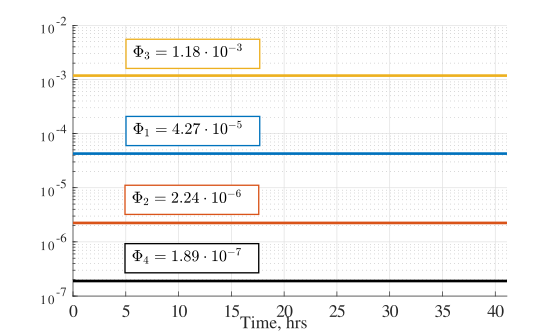}
    \caption{Conserved quantities for the short time LEO to GEO transfer}
    \label{conservationLaws3Dplot}
\end{figure}

The short transfer time trajectory is plotted in Fig. \ref{shorttrajectory}, with a black dashed line representing the target orbit. Over the span of just under two days, the spacecraft is transferred from an inclined LEO to an equatorial GEO orbit. The time history of the control accelerations can be seen in Fig. \ref{shorttrajectorycontrol}. 
The functions $\Phi_1$ to $\Phi_4$ in Eq. \eqref{conservationLaws3D} were evaluated along the state and control time histories for this transfer. The resulting values were plotted over the transfer time in Fig. \ref{conservationLaws3Dplot}. Note that as the values differ largely in magnitude, the $\Phi_i$ were plotted on a logarithmic scale to visually separate them. Curves therefore represent the absolute values of the conserved quantities. The signed value of each $\Phi$ is given over each line. 
\noindent The figure demonstrates that the functions $\Phi_1$ to $\Phi_4$ are invariants over the state and control trajectories for the optimal control system. These invariants span many orders of magnitude, but individually each does not vary by more than around $10^{-14}$. The variations are caused by the effects of floating point precision in the integration of the equations of motion. 

For the second transfer, the same trajectory was solved using a significantly longer transfer time, resulting in a 190 revolution spiral out from 28$^\circ$ inclined LEO to an equatorial GEO. {The initial conditions of the transfer are shown in Table \ref{ICs2}, with a transfer time of $5000 \,\mathrm{TU}$.} 

\begin{table}[hbt!]
\caption{\label{ICs2} Initial conditions for long transfer in canonical units, $t_f=5000\,\mathrm{TU}$}
\centering
\begin{tabular}{crcr}
\hline
State & Initial Value & Control & Initial Value \\\hline
$x$ & $-0.24219$ & $u_x$ & $2.73691\cdot 10^{-4}$ \\
$y$ & $-1.01025$ & $u_y$ & $-7.71318\cdot 10^{-5}$\\ 
$z$ & $0.13036$ & $u_z$ & $-8.83630\cdot 10^{-5}$\\ 
$\dot{x}$ & $0.83402$ & $\dot{u}_x$ & $6.93387\cdot10^{-5}$ \\ 
$\dot{y}$ & $-0.25776$ & $\dot{u}_y$ & $2.63684\cdot 10^{-4}$ \\
$\dot{z}$ & $-0.44305$ & $\dot{u}_z$ & $-2.27411\cdot 10^{-5}$ \\
\hline
\end{tabular}
\end{table}

\begin{figure}[hbt!]
    \centering
    \includegraphics[width=1\textwidth]{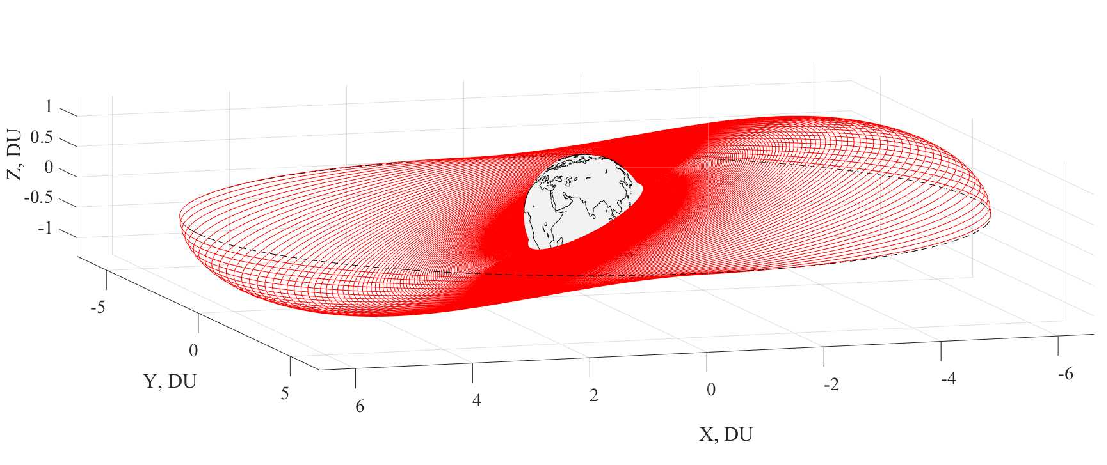}
    \caption{Optimal LEO to GEO transfer, $\Delta t\approx$ 46 days}
    \label{longtrajectory}
\end{figure}

\begin{figure}[hbt!]
    \centering
    \includegraphics[width=0.9\textwidth]{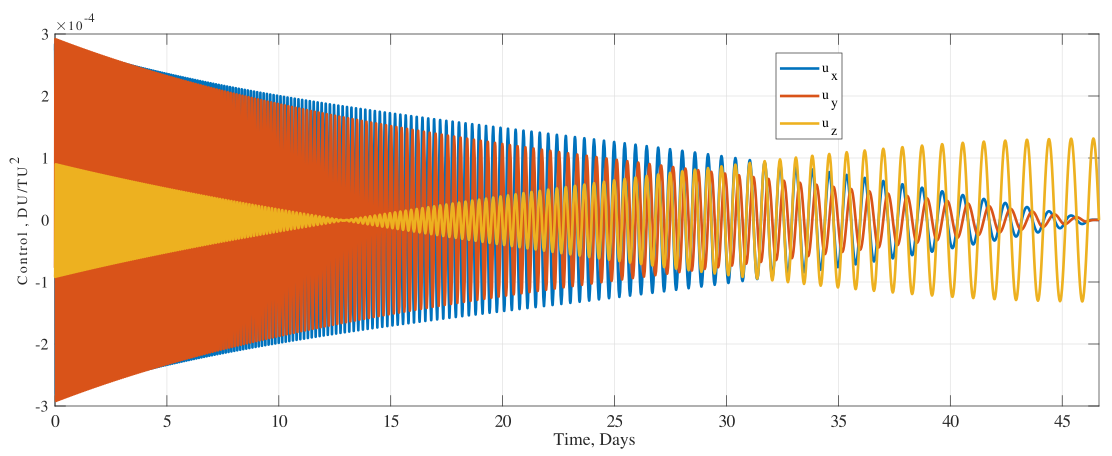}
    \caption{Control time history for the long time LEO to GEO transfer}
    \label{longtrajectorycontrol}
\end{figure}

\begin{figure}[hbt!]
    \centering
    \includegraphics[width=0.8\textwidth]{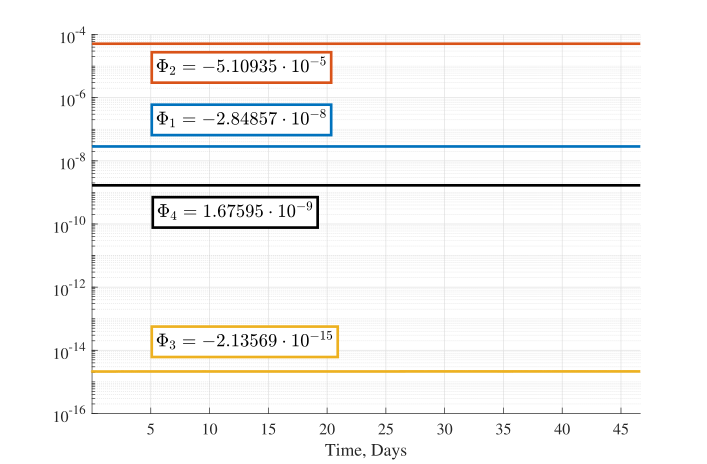}
    \caption{Conserved quantities for the long time LEO to GEO transfer}
    \label{longconservationLaws3D}
\end{figure}

The trajectory for this transfer can be seen in Fig. \ref{longtrajectory}. As before, the black dashed line represents the target orbit. The time histories of the three control components can be seen in Fig. \ref{longtrajectorycontrol}. The maximum magnitude for the controls is two orders of magnitude smaller in this case than the short transfer time case. 
\noindent The time histories of $\Phi_1$ to $\Phi_4$ were once again evaluated along the state and control paths and plotted in Fig. \ref{longconservationLaws3D}. For this transfer, several of the integrals of motion were negative; this is clearly indicated for the curves when applicable.

 As before, it is clear from the figure that the four conserved quantities are in fact constant over the trajectories of the system. Similarly to before, the conserved quantities span many orders of magnitude; the smallest on the order of magnitude of $10^{-15}$. To properly capture the invariance of $\Phi_3$, very tight integrator tolerances were required. What this may indicate is that for very long spiraling transfers such as the one presented here, non-canonical units may be a better choice for the integration when computing path invariances due to the larger magnitudes of the numbers involved.

\section{Conclusion}
By applying the Lie theory of infinitesimal transformations to the generalized Lagrangian approach to optimal control, a new set of conserved quantities was found for the minimum energy continuous thrust optimal control problem in an inverse squared gravitational field. By using the fact that the optimal control equations can be derived from an action principle, the Killing equations were formulated and solved in Cartesian coordinates to arrive at a set of infinitesimal coordinate and time transformations which leave the action functional invariant. After applying these infinitesimal transformations to Noether's theorem, four linearly independent conserved quantities were derived in the Cartesian optimal control problem. It was then demonstrated how to transform these conserved quantities between coordinate frames, by finding equivalent infinitesimal transformations in the new coordinates and applying Noether's theorem to those. To demonstrate the technique, the Cartesian conserved quantities were transformed first into a polar and then into a spherical coordinate frame. The invariance of these conserved quantities was demonstrated numerically through two sample optimal continuous thrust transfers with a varying number of revolutions. 

\section*{Acknowledgments}

This material is based upon work supported by the National Science Foundation Graduate Research Fellowship Program under Grant No. 2336877.

\section*{Appendix}

{
\subsection{Generalized Lagrangian: Proof of Optimal Control Equations}
Equations \eqref{cartesiangeneralizedLagrangian} and \eqref{polargeneralizedLagrangian} are two generalized Lagrangians for the minimum energy optimal control problem in Cartesian and polar coordinates, respectively. A proof is presented here for the form of the Cartesian generalized Lagrangian; the polar generalized Lagrangian has a similar proof which follows the one presented here. Consider the minimum energy optimal control problem, with continuous cost $$\mathrm{min\,}\mathcal{J}=\frac{1}{2}\int_{t_0}^{t_f}\bm{u}^T\bm{u}\,dt$$ Following the classical optimal control theory, the Hamiltonian function is constructed by appending the differential equation constraints to the continuous cost of the system. $$H = \frac{1}{2}\bm{u}^T\bm{u}+\bm{\lambda}_r\bm{v}+\bm{\lambda}_v\left(-\frac{\mu}{r^3}\bm{r}+\bm{u}\right)$$ Applying the necessary conditions for optimality, the costate differential equations can be written as 
\begin{align}
    \dot{\bm{\lambda}}^T_r&=-\p{H}{\bm{r}} = \mu\bm{\lambda}_v^T\left(\frac{I_{3\times 3}}{r^3} - \frac{3\bm{r}\bm{r}^T}{r^5}\right ) \\ 
    \dot{\bm{\lambda}}^T_v &=-\p{H}{\bm{v}} = -\bm{\lambda}_r^T
\end{align}
Applying the PMP to the Hamiltonian yields the optimal control law $\bm{u}=-\bm\lambda_v$, this result following directly from the primer vector theory \cite{conway2010spacecraft}. This optimal control law can be combined with the second costate differential equation ($\bm{\lambda_r}=-\dot{\bm{\lambda}}_v=\dot{\bm{u}}$) to rewrite the costate differential equations into a second order  differential equation in the control rather than the costates. 
\begin{equation}
    \ddot{u} = -\mu\bm{u}^T\left(\frac{I_{3\times 3}}{r^3} - \frac{3\bm{r}\bm{r}^T}{r^5}\right )
\end{equation}
Expanding this equation out into scalar form yields the three scalar differential equations for the control components. 
\begin{equation}
    \begin{split}
        \ddot{u}_x &= -\frac{\mu}{r^3}u_x + \frac{3\mu}{r^5}\left(x^2u_x + xyu_y + xzu_z \right) \\ 
        \ddot{u}_y &= -\frac{\mu}{r^3}u_y + \frac{3\mu}{r^5}\left( y^2u_y + xyu_x + yzu_z\right) \\
        \ddot{u}_z  &= -\frac{\mu}{r^3}u_z + \frac{3\mu}{r^5}\left(z^2u_z + xzu_x + yzu_y \right)
    \end{split}
\end{equation}
These are the exact same differential equations for the control as those arrived at by using the Cartesian generalized Lagrangian, shown in Eq. \eqref{cartesianEOMs}. The generalized Lagrangian therefore leads to the correct state and control differential equations. A similar proof can be constructed to show the correctness of the polar generalized Lagrangian. 
}

\subsection{Optimal Control Equations for Spherical Two Body Problem}
Minimizing the modified action functional given by the time integration of Eq. \eqref{sphericalgeneralizedLagrangian} with respect to $\tilde{\bm{q}}=[r,\theta,\phi,u_r,u_\theta,u_\phi]^T $yields the following state and control differential equations.

\begin{align}
    \ddot{r} &= u_r+r{\dot{\phi}}^2 -\frac{\mu }{r^2}+r {\dot{\theta}}^2 {\cos{\phi}}^2\\
    \ddot{\theta} &= u_\theta+2 \dot{\theta} \dot{\phi} \tan{\phi}-\frac{2 \dot{r} \dot{\theta}}{r}\\
    \ddot{\phi} &= u_\phi-\frac{2 \dot{r}\dot{\phi} }{r}-{\dot{\theta}}^2 \cos{\phi} \sin{\phi}\\ 
    \ddot{u}_r &= \left({\dot{\phi}}^2 +{\dot{\theta}}^2 \cos^2{\phi}+\frac{2 \mu }{r^3} \right)u_r+\frac{2 \ddot{\theta}}{r} u_\theta+\frac{2 \ddot{\phi}}{r} u_\phi+\frac{2}{r}\left(\dot{\phi}  \dot{u}_\phi+ \dot{\theta} \dot{u}_\theta\right)\\ 
    \ddot{u}_\theta &=
    \begin{array}[t]{ll}
     &\left(2  r \dot{\theta} \dot{\phi}\sin{2\phi}-2 \dot{r} \dot{\theta} \cos^2{\phi}-2 r \ddot{\theta} \cos^2{\phi}\right) u_r+\left(\frac{2 \ddot{r}}{r}-2 {\dot{\phi}}^2 {\tan^2{\phi}}-\frac{2 {\dot{r}}^2}{r^2}-2 \ddot{\phi} \tan{\phi}-2 {\dot{\phi}}^2\right) u_\theta \\
     &+\left(2 \dot{\theta} \dot{\phi}  \cos{2\phi}+\ddot{\theta} \sin{2\phi}\right) u_\phi+\frac{2 \dot{r} \dot{u}_\theta}{r}-2 \dot{\phi} \dot{u}_\theta \tan{\phi}+\dot{\theta} \dot{u}_\phi \sin{2\phi}-2 r \dot{\theta} \dot{u}_r \cos^2{\phi}\\ 
    \end{array}\\
    \ddot{u}_\phi &= 
    \begin{array}[t]{ll}
    &\left(-r {\dot{\theta}}^2\sin{2\phi}-2 \ddot{\phi} r-2 \dot{\phi} \dot{r}\right) u_r-{2 \ddot{\theta} \tan{\phi}} u_\theta+\left(\frac{2 \ddot{r}}{r}-{\dot{\theta}}^2 \left(2 \cos^2{\phi}-1\right)-\frac{2 {\dot{r}}^2}{r^2}\right) u_\phi \\ 
    &+\frac{2 \dot{r} \dot{u}_\phi}{r}-2 r\dot{\phi}  \dot{u}_r-(2 \dot{\theta} \tan{\phi})\dot{u}_\theta  
    \end{array}
\end{align}

\subsection{Killing Equations for Three Dimensional Case}
The Killing equations are the set of partial differential equations which must be satisfied by the generators to achieve invariance of the functional under the infinitesimal transformation. There are 46 Killing equations for the three dimensional two body problem which the generators must satisfy. Of these, 24 equations state what variables each generator is not a function of. 
\begin{equation}
    \begin{split}
        \p{\eta_1}{u_x}=\p{\eta_2}{u_y}=\p{\eta_3}{u_z}=\p{\eta_4}{x}=\p{\eta_5}{y}=\p{\eta_6}{z}&=0 \\ 
        \p{\xi}{x}=\p{\xi}{y}=\p{\xi}{z}=\p{\xi}{u_x}=\p{\xi}{u_y}=\p{\xi}{u_z}&=0
    \end{split}
\end{equation}
\noindent The second condition states that $\xi(t,x,y,u_x,u_y,u_z)=\xi(t)$. As in the 2D case, we let $\dot{\xi}=f(t)$. Of the remaining 22 Killing equations, fifteen do not include the divergence term $\varphi$. These equations are

\begin{table}[hbt!]
\centering
\begin{tabular}{lll}

\parbox{5cm}{
\begin{equation}
\p{\eta_4}{u_x}+\p{\eta_1}{x}= f 
\end{equation}}

& 
\parbox{5cm}{
\begin{equation}
\p{\eta_5}{u_y}+\p{\eta_2}{y}= f 
\end{equation}}       

& 
\parbox{5cm}{
\begin{equation}
    \p{\eta_6}{u_z}+\p{\eta_3}{z}= f 
\end{equation}}   

\\

\parbox{5cm}{
\begin{equation}
\p{\eta_2}{u_x}+\p{\eta_1}{u_y}=0 
\end{equation}}

& 
\parbox{5cm}{
\begin{equation}
\p{\eta_3}{u_x}+\p{\eta_1}{u_z}=0 
\end{equation}}       

& 
\parbox{5cm}{
\begin{equation}
\p{\eta_3}{u_y}+\p{\eta_2}{u_z}=0 
\end{equation}}   

\\

\parbox{5cm}{
\begin{equation}
\p{\eta_5}{x}+\p{\eta_4}{y}=0 
\end{equation}}

& 
\parbox{5cm}{
\begin{equation}
\p{\eta_6}{x}+\p{\eta_4}{z}=0 
\end{equation}}       

& 
\parbox{5cm}{
\begin{equation}
\p{\eta_6}{y}+\p{\eta_5}{z}=0 
\end{equation}}   

\\

\parbox{5cm}{
\begin{equation}
\p{\eta_4}{u_y}+\p{\eta_2}{x}=0 
\end{equation}}

& 
\parbox{5cm}{
\begin{equation}
\p{\eta_5}{u_x}+\p{\eta_1}{y}=0 
\end{equation}}       

& 
\parbox{5cm}{
\begin{equation}
\p{\eta_6}{u_x}+\p{\eta_1}{z}=0 
\end{equation}}   

\\

\parbox{5cm}{
\begin{equation}
\p{\eta_4}{u_z}+\p{\eta_3}{x}=0 
\end{equation}}

& 
\parbox{5cm}{
\begin{equation}
\p{\eta_5}{u_z}+\p{\eta_3}{y}=0 
\end{equation}}       

& 
\parbox{5cm}{
\begin{equation}
\p{\eta_6}{u_y}+\p{\eta_2}{z}=0 
\end{equation}}   

\\

\end{tabular}
\end{table}

\noindent The solution to the preceding 39 PDEs is computed in the same way as was done in the two dimensional case, as all of these equations eventually reduce to a linearity condition of a certain variable in each generator. The solution is of the form 
\begin{equation}
\begin{split}
    \dot\xi(t)&=f \\ 
    \eta_1(t,x,y,z,u_y,u_z) &= (f-c_4)x - c_9y - c_{14}z - c_1u_y-c_2u_z+h_1 \\
    \eta_2(t,x,y,z,u_x,u_z) &= -c_5x + (f-c_{10})y - c_{15}z + c_1u_x - c_3u_z + h_2 \\
    \eta_3(t,x,y,z,u_x,u_y) &= -c_6x-c_{11}y + (f-c_7)z + c_2u_x+c_3u_y+h_3 \\
    \eta_4(t,y,z,u_x,u_y,u_z)&= -c_8y-c_{12}z+c_4u_x+c_5u_y+c_6u_z+h_4\\
    \eta_5(t,x,z,u_x,u_y,u_z)&= c_8 x - c_{13}z + c_9u_x + c_{10}u_y + c_{11}u_z + h_5 \\ 
    \eta_6(t,x,y,u_x,u_y,u_z)&= c_{12}x+c_{13}y+c_{14}u_x+c_{15}u_y+c_7u_z+h_6 
\end{split}
\end{equation}
\noindent All coefficients $c_1$ to $c_{15}$ as well as $h_1$ to $h_6$ are functions of time. The remaining seven Killing equations deal with the divergence term, and relate time derivatives of each of the generators to partial derivatives of $\varphi$ with respect to each of the coordinates and time. Let $\bm{\eta}_r=[\eta_1, \eta_2, \eta_3]^T$ and $\bm{\eta}_u=[\eta_4, \eta_5, \eta_5]^T$. The remaining Killing equations which must be satisfied by the generators are 

\begin{table}[hbt!]
\begin{tabular}{ll}

\parbox{7cm}{
\begin{equation}
    \p{\eta_1}{t}   = \p{\varphi}{u_x} 
\end{equation}}

& 
\parbox{7cm}{
\begin{equation}
        \p{\eta_2}{t} = \p{\varphi}{u_y} 
\end{equation}}       

\\

\parbox{7cm}{
\begin{equation}
\p{\eta_3}{t} = \p{\varphi}{u_z} 
\end{equation}}

& 
\parbox{7cm}{
\begin{equation}
        \p{\eta_4}{t} = \p{\varphi}{x} 
\end{equation}}       

\\

\parbox{7cm}{
\begin{equation}
        \p{\eta_5}{t} = \p{\varphi}{y} 
\end{equation}}

& 
\parbox{7cm}{
\begin{equation}
        \p{\eta_6}{t} = \p{\varphi}{z} 
\end{equation}}       

\\

\multicolumn{2}{l}
{
\parbox{0.95\textwidth}{
\begin{equation}
\p{\varphi}{t} = \frac{3\mu\bm{u}^T\bm{r}}{r^5}\bm{r}^T\bm{\eta}_r - \frac{\mu}{r^3}(\bm{u}^T\bm{\eta}_r+\bm{r}^T\bm{\eta}_u+f\bm{u}^T\bm{r}) + \bm{u}^T\bm{\eta}_u+\frac{f}{2}\bm{u}^T\bm{u}
\end{equation} }
}
\end{tabular}
\end{table}

The general solution to the 3D set of Killing equations is 
\begin{equation}
\begin{split}
    \xi &= k_\xi \\ 
    \eta_1(t,x,y,z,u_y,u_z) &= - k_9y  + k_{6}z  \\
    \eta_2(t,x,y,z,u_x,u_z) &=   k_9x  + k_{11}z  \\
    \eta_3(t,x,y,z,u_x,u_y) &=  -k_6x   -k_{11}y   \\
    \eta_4(t,y,z,u_x,u_y,u_z)&= -k_9u_y + k_6u_z\\
    \eta_5(t,x,z,u_x,u_y,u_z)&=  k_9u_x + k_{11}u_z\\ 
    \eta_6(t,x,y,u_x,u_y,u_z)&= -k_{6}u_x-k_{11}u_y
\end{split}
\end{equation}
\noindent The four constants, $k_\xi,k_6,k_9,k_{11}$ define four linearly independent solutions to the Killing equations, and as such the four linearly independent conserved quantities are found by applying each of these to Noether's theorem. 

\bibliographystyle{unsrt} 
\bibliography{sample,ORefs,myRefs}

@ARTICLE{11,
  AUTHOR =       {Ely, T. A., Crossley, W. A., and Williams, E. A.},
  TITLE =        {Satellite Constellation Design for Zonal Coverage Using Genetic Algorithms},
 JOURNAL =      {Journal of the Astronautical Sciences},
  YEAR =         {Jul-Dec 1999},
  volume =       {47},
  number =       {3-4},
  pages =        {207--228},
  doi =          {doi: 10.2514/1.4191}
}

@book{conway2010spacecraft,
  title={Spacecraft trajectory optimization},
  author={Conway, B.},
  volume={29},
  year={2010},
  publisher={Cambridge Univ Press},
doi = "10.1017/cbo9780511778025 "
}

@BOOK{Bryson1975,
author = {Jr. Arthur E. Bryson  and Yu-Chi Ho},
title = {Applied Optimal Control optimization,estimation and control},
publisher = {CRC Press,},
year = {1975},
doi = "10.1201/9781315137667"
}

@article{rao2009survey,
  title={A survey of numerical methods for optimal control},
  author={Rao, Anil V},
  journal={Advances in the Astronautical Sciences},
  volume={135},
  number={1},
  pages={497--528},
  year={2009},
  publisher={Univelt, Inc.}
}

@TECHREPORT{3,
  author = "Navagh, J.",
  title =        {Optimizing Interplanetary Trajectories with Deep Space Maneuvers},
  institution =  {NASA Langley Research Center},
  year =         {1993},
  type =         {NASA Contractor Report 4546},
  address =      {Hampton, VA},
}

@article{15,
title = "Preliminary Design of Multiple Gravity-Assist Trajectories",
journal = "Journal of Spacecraft and Rockets",
volume = "43",
number = "4",
pages = " 794--805",
year = "Jul.-Aug. 2006",
author = "Vasile, M. and Pascale, P. D.",
}

@INPROCEEDINGS{OCP2025AAShawaii,
  AUTHOR =      {Aimar Negrete and Ossama Abdelkhalik},
  TITLE =       {Variational Methods Of Analytical Mechanics For Trajectory Optimization},
  BOOKTITLE =     {AAS/AIAA Space Flight Mechanics Meeting},
  number =      {AAS 25-416},
  month =       {January 19-23},
  year =        {2025},
  address =     {Kaua'i, Hawaii},
}

@article{lutzky1978symmetry,
  title={Symmetry groups and conserved quantities for the harmonic oscillator},
  author={Lutzky, M},
  journal={Journal of Physics A: Mathematical and General},
  volume={11},
  number={2},
  pages={249},
  year={1978},
  publisher={IOP Publishing},
doi="10.1088/0305-4470/11/2/005 "
}

@article{vujanovic1970group,
  title={A group-variational procedure for finding first integrals of dynamical systems},
  author={Vujanovic, B},
  journal={International Journal of Non-Linear Mechanics},
  volume={5},
  number={2},
  pages={269--278},
  year={1970},
  publisher={Elsevier},
doi="10.1016/0020-7462(70)90024-7 "
}

@BOOK{neuenschwander,
  author =       {D. Neuenschwander},
  title =        {Emmy Noether's Wonderful Theorem},
  year =         {2017},
  publisher =    {Johns Hopkins University Press},
doi = "10.56021/9781421422671 "
}

@incollection{noether1983invariante,
  title={Invariante variationsprobleme},
  author={Noether, Emmy},
  booktitle={Gesammelte Abhandlungen-Collected Papers},
  pages={231--239},
  year={1983},
  publisher={Springer},
doi="10.1007/978-3-642-61761-4_15 "
}

@article{Noether1971,
   title={Invariant variation problems, English Translation},
   volume={1},
   ISSN={1532-2424},
   DOI={10.1080/00411457108231446},
   number={3},
   journal={Transport Theory and Statistical Physics},
   publisher={Informa UK Limited},
   author={Noether, Emmy},
   year={1971},
   month=jan, pages={186–207} }

@book{cohen1911introduction,
  title={An Introduction to the Lie Theory of one-parameter groups: With Applications to the solution of Differential Equations},
  author={Cohen, Abraham},
  year={1911},
  publisher={DC Health \& Company},
doi="10.1126/science.34.887.924 "
}

@book{olver1993applications,
  title={Applications of Lie Groups to Differential Equations},
  author={Olver, P},
  year={1993},
  publisher={Springer-Verlag},
doi="10.1007/978-1-4612-4350-2 "
}

@book{wiener1949,
  author    = {Norbert Wiener},
  title     = {Extrapolation, Interpolation, and Smoothing of Stationary Time Series},
  publisher = {MIT Press},
  year      = {1949},
  doi = "10.7551/mitpress/2946.001.0001 "
}

@book{bellman1957,
  author    = {Richard Bellman},
  title     = {Dynamic Programming},
  publisher = {Princeton University Press},
  year      = {1957},
doi = "10.1515/9781400835386 "
}

@book{pontryagin1962,
  author    = {Lev S. Pontryagin and V. G. Boltyanskii and R. V. Gamkrelidze and E. F. Mishchenko},
  title     = {The Mathematical Theory of Optimal Processes},
  publisher = {Interscience Publishers},
  year      = {1962},
doi = "10.1201/9780203749319"

}

@book{stoer2002introduction,
  author    = {Stoer, Josef and Bulirsch, Roland},
  title     = {Introduction to Numerical Analysis},
  publisher = {Springer},
  year      = {2002},
  edition   = {3rd},
  isbn      = {978-0387954523},
doi="10.1007/978-1-4757-5592-3 "
}

@book{betts2010practical,
  author    = {Betts, John T.},
  title     = {Practical Methods for Optimal Control and Estimation Using Nonlinear Programming},
  publisher = {SIAM},
  year      = {2010},
  edition   = {2nd},
  isbn      = {978-0898718577},
doi="10.1137/1.9780898718577 "
}

@article{betts1998survey,
  author = {Betts, John T.},
  title = {Survey of numerical methods for trajectory optimization},
  journal = {Journal of Guidance, Control, and Dynamics},
  volume = {21},
  number = {2},
  pages = {193--207},
  year = {1998},
doi="10.2514/2.4231 "
}
\end{document}